\documentclass[12pt]{iopart}

\usepackage{amssymb}
\usepackage{amsthm}
\usepackage{graphicx}
\usepackage{url}

\newcommand{\Ker}{\mathop{\rm Ker}}
\newcommand{\Range}{\mathop{\rm Range}}

\newtheorem{lemma}{Lemma}[section]
\newtheorem{theo}{Theorem}[section]
\newtheorem{cor}[theo]{Corollary}
\newtheorem{prop}{Proposition}
\newtheorem{conj}{Conjecture}

\theoremstyle{remark}
\newtheorem{rem}{Remark}

\newcommand{\R}{\mathbb R}

\begin{document}

\title{Multiplicity of periodic solutions in bistable equations}
\author{Gregory Berkolaiko\footnote{Present address: Mathematics,
    Texas A\&M University, College Station, TX 77843-3368, USA} and
  Michael Grinfeld} 
\address{
  Department of Mathematics,\\
  University of Strathclyde,\\
  Livingstone Tower, 26 Richmond Street,\\
  Glasgow G1 1XH, Scotland, U.K.}
\ead{Gregory.Berkolaiko@math.tamu.edu}

\begin{abstract}
  We study the number of periodic solutions in two first order
  non-autonomous differential equations both of which have been used to
  describe, among other things, the mean magnetization of an Ising
  magnet in the time-varying external magnetic field.  When the
  strength of the external field is varied, the set of periodic solutions
  undergoes a bifurcation in both equations.  We prove that despite
  profound similarities between the equations, the character of the
  bifurcation can be very different.  This results in a different
  number of coexisting stable periodic solutions in the vicinity of
  the bifurcation.  As a consequence, in one of the models, the
  Suzuki-Kubo equation, one can effect a discontinuous change in
  magnetization by adiabatically varying the strength of the magnetic
  field.
\end{abstract}

\pacno{05.50.+q}

%%%%%%%%%%%%%%%%%%%%%%%%%%%%%%%%%%%%%%%%%%%%%%%%%%%%%%%%%%%%%%%%%%%%%%%%%
%%%%%%%%%%%%%%%%%%%%%%% Section %%%%%%%%%%%%%%%%%%%%%%%%%%%%%%%%%%%%%%%%%
\section{Introduction}

In many lattice models the mean field approximation leads to ODEs with
periodically time dependent right-hand side. For example, the
Curie-Weiss model with Glauber spin dynamics in the thermodynamic
limit leads to the so-called Suzuki-Kubo equation (see the derivation
in \cite[Ch. 3]{Berg} or in \cite{Rao,SK}, and a numerical study in
\cite{AcC}).

The Suzuki-Kubo equation is
\begin{equation} \label{sk}
  \epsilon \frac{dm}{d\tau} = -m 
  + \tanh [ \beta (m + \tilde{h}\cos(2\pi\tau))],
\end{equation}
where $m$ is the average magnetization of a sample, $\epsilon$ is the
relaxation time, $\tilde{h}$ is the amplitude of the applied magnetic
field, and $\beta$ is $1/kT$, with $T$ the absolute temperature.

A seemingly similar first order non-autonomous equation,
\begin{equation}  \label{ob}
  \epsilon\frac{dx}{dt} =  a x + b x^3 +  h\cos (2\pi t), 
  \; a, \, b \in  \R,
\end{equation}
has been used as a simple generic model for switchable bistable systems.
If $a>0$ and $b<0$, it describes the overdamped dynamics of a particle
in a quadratic potential with periodic forcing.  Such equations have
been studied in the context of laser optics (longitudinal mode
instabilities in a semiconductor laser); see \cite{Hohl,JGR} for
analysis and references to the optics literature.

%%%%% something about the variables %%%%%%%%

The issue we want to address is the number of $2\pi$-periodic
solutions for various values of $h$ and other parameters.
In particular, we will use $h$ as the bifurcation parameter. This
issue is of physical importance, since one wants to determine whether
it is possible, say, in the Suzuki-Kubo context, to effect a
discontinuous change in magnetization by continuously varying the
amplitude of the applied magnetic field $h$ (i.e. a first order phase
transition).

To compare the two equations effectively, we rescale the variables in
(\ref{sk}) in the following way: set, in sequence,
\begin{equation}
  \fl
  x = m +\tilde{h}\cos(2\pi\tau), \quad
  h = \tilde{h}\sqrt{1+4\pi^2\epsilon^2}, \quad
  t = \tau + \frac{1}{2\pi}
  \arccos\left(\frac{1}{\sqrt{1+4\pi^2\epsilon^2}}\right).
\end{equation}
Then in the new variables we have 
\begin{equation}\label{nsk}
  \epsilon\frac{dx}{dt} = -x + \tanh \beta x +h \cos(2\pi t).
\end{equation}

Comparing the two equations, we see that the nonlinearity in
(\ref{nsk}) is sublinear while it is superlinear in (\ref{ob}).
Furthermore, $\tanh(z)$ is not an entire function.  Note also that
the Taylor series expansion of $-x+\tanh \beta x$ is
\begin{equation}
  -x+\tanh \beta x = (\beta-1) x - \frac{\beta^3}{3}x^3 + O(x^5).
\end{equation}
Hence superficially it looks that the behaviour of solutions of
(\ref{nsk}) should be similar to that of (\ref{ob}) with $a=\beta-1$
and $b=-\beta^3$.  In fact, this is the implicit assumption under much
of the physical literature.  This assumption is incorrect.  It turns
out that for certain values of $\beta$ and $h$ equation (\ref{nsk})
has 5 periodic solution, a situation which is impossible for equation
(\ref{ob}).  The numerical simulations confirming this were reported
in, for example, \cite[Ch. 7]{Berg}.  Here we aim to establish this
fact analytically.

The structure of the paper is as follows.  After collecting
preliminaries below, in section~\ref{LS} we perform the
Liapunov-Schmidt reduction of the general equation,
\begin{equation} \label{ge}
  \frac{dx}{dt}= f(x) + h\cos(2\pi t),
\end{equation}
The reduction is complicated by the fact that we do not know the
bifurcating solution explicitly.  

In section~\ref{inf} we use the results of the reduction to conclude
that for small $\beta$ behaviour of (\ref{nsk}) is similar to that of
(\ref{ob}).  Then we consider the case of large $\beta$ in (\ref{nsk})
by explicitly examining the case of $\beta=\infty$.  The limiting
equation is in fact a differential inclusion which we analyse explicitly,
finding the range of $h$ for which 5 solutions co-exist.  Appealing to
a continuation theorem \cite{AC}, we conclude that such behaviour
persist for large, but finite, values of $\beta$ as well.

Everywhere below we assume that the function $f(\cdot)$ is
sufficiently smooth.  As can be seen from (\ref{ge}), we fix $\epsilon
= 1$, as this does not affect any of the results.  However in our
numerics we take $\epsilon$ of the order of $0.05$ which ``slows
down'' the time and results in nicer plots.

%============ The hysteresis loop (cubic, e=1/20, h=2.0) ====
\begin{figure}[t]
  \centerline{
    \includegraphics[height=8cm,width=12cm]{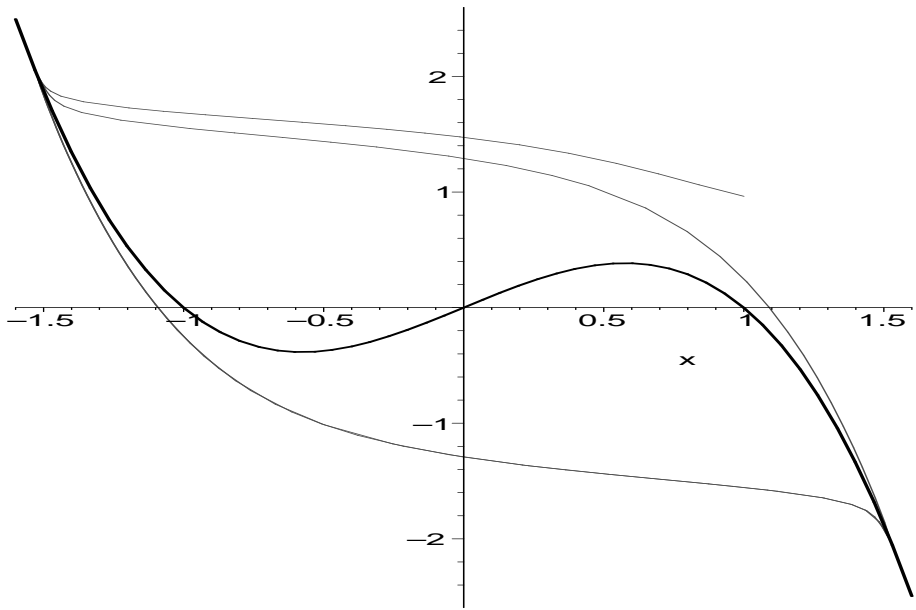}
  }
  \caption{The picture of the (stable) hysteresis loop for large $h$. 
    In this case we take $f(x) = x - x^3$, $\epsilon = 0.05$ and $h =
    2.0$.}
  \label{fig:hyst_loop}
\end{figure}
%============================================================

Whenever we sketch the solutions of the equation we do it in
$(x, h\cos(2\pi t))$ plane, i.e.~we plot the curve $(x(t), y(t))$ where
\begin{eqnarray}
  \label{eq:phase_space}
  \cases{\dot{x} = f(x) - y,&\\
    y = h\cos(2\pi t).&}
\end{eqnarray}
This makes it easier to identify the periodic solutions (they are
closed curves) and produces familiar hysteresis loop pictures for large
$h$, figure~\ref{fig:hyst_loop}.

%%%%%%%%%%%%%%%%%%%%%%%%%%%%%%%%%%%%%%%%%%%%%%%%%%%%%%%%%%%%%%%%%%%%%%%%%
%%%%%%%%%%%%%%%%%%%%%%% Section %%%%%%%%%%%%%%%%%%%%%%%%%%%%%%%%%%%%%%%%%
\section{Background information}

The examples of function $f(x)$ that we are going to consider here,
$f(x) = \tanh\beta x - x$, $\beta > 1$ and $f(x) = \gamma x - x^3$,
$\gamma > 0$ share many important properties.  Both are odd functions
and the equation $f(x)=0$ has exactly three solutions.  If we consider
equation~(\ref{ge}) for $h=0$, there will correspondingly be three
stationary solutions.  The central stationary solution $x(t) = 0$ will
be unstable and the other two are locally asymptotically stable.

\begin{figure}[t]
  \centerline{\includegraphics[scale=0.7]{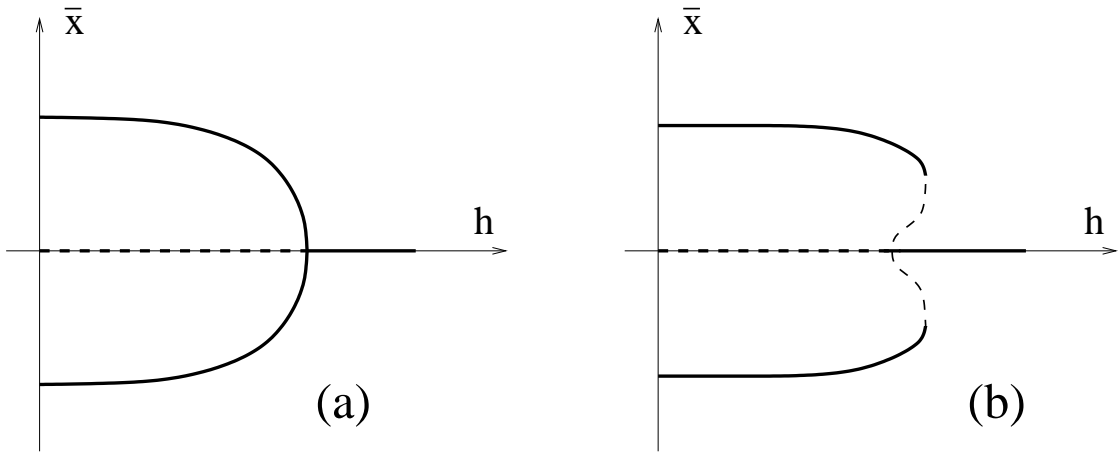}}
  \caption{Two possible minimal scenarios of the emergence of a single
    periodic solution.  In scenario (a) three solutions merge in a
    subcritical pitchfork bifurcation.  In scenario (b) the central
    solution first undergoes a supercritical pitchfork bifurcation,
    emitting two unstable solutions.  These unstable solutions
    disappear in fold bifurcations upon meeting the stable solutions.}
  \label{fig:fork}
\end{figure}

It follows from the implicit function theorem that for small $h$,
there will be three periodic solutions, two stable and one unstable.
Below we show that this situation persists at least until $h_0 =
\max_{x>0} f(x)$.  Also we will show that for large $h$ there is only
one, stable, solution.  The main aim of this paper is to uncover the
bifurcation picture for the intermediate values of $h$.  The simplest
scenario where three solutions can merge into one is a subcritical
pitchfork as shown on figure \ref{fig:fork}(a).  There, for
simplicity, we sketch the means of periodic solutions, $\bar{x} =
\int_0^1 x(t) dt$, versus $h$.

%=== Sketch of an orbit crossing [-x_1, x_1] ============
%=========================================================

Contrary to the above expectations, numerical investigations of the
Suzuki-Kubo equation~(\ref{nsk}) performed in \cite{TO} (see also
\cite[Ch. 3]{Berg} and \cite{AC}) show that for large values of
$\beta$ and the parameter $h$, the equation has {\sl three} stable
periodic solutions (and therefore five periodic solutions altogether).
The bifurcation diagram would then look like
figure \ref{fig:fork}(b).  For small values of $\beta$, \cite{TO}
reported two stable periodic solutions at most, a situation compatible
with figure \ref{fig:fork}(a).  The critical value of $\beta$ at
which the bifurcation picture changed was referred to as the {\sl
  tricritical points} (TCP).

The existence of TCP was disputed by \cite{Zimmer} (see also
\cite{KRN}), who studied equation~(\ref{ob}) numerically and reported
that, although the stable non-symmetric solution changed fast in the
vicinity of bifurcation, it was not disappearing in a ``blue-sky
catastrophe'' as in figure \ref{fig:fork}(b).

Our investigation concludes that, from mathematical point of view,
both results are correct: there are at most 3 solutions in (\ref{ob})
and there is a TCP in equation (\ref{nsk}).

%%%%%%%%%%%%%%%%%%%%%%% Section %%%%%%%%%%%%%%%%%%%%%%%%%%%%%%%%%%%%%%%%%
\subsection{Periodic solutions for small $h$}

\begin{theo} \label{small_h}
  Let $f(x)$ be a continuously differentiable function satisfying
  \begin{equation}
    \label{eq:small_h_cond}
    f'(x)<0 \quad x\in(a,b)
    \quad\mbox{and}\quad f(a) > 0 > f(b).
  \end{equation}
  Put $h_0 = \min(f(a), -f(b))$.  Then for any $h < h_0$ there is a
  unique periodic solution to equation (\ref{ge}) satisfying
  $x(t)\in[a,b]$ for all $t$.  This solution is stable.
\end{theo}

\begin{proof}
  The idea of the proof is to construct such a trapping region that any
  solution that enters it must stay there.  Then we can appeal to a
  fixed point theorem to infer that there is a periodic solution
  inside this region.  Analysing stability of this solution we find
  that it must be stable, therefore it is unique.
  
  The function $f(x)$ is monotone therefore there are unique solutions
  to the equations $f(x_a)=h_0$ and $f(x_b)=-h_0$.  Since, for any $t$,
  $x(t)=x_a$ implies $x'(t) = f(x_a) + h\cos(2\pi t) \geq h_0 - h > 0$,
  any solution to the right of $x_a$ will remain there.  Similarly,
  solutions to the left of $x_b$ will never increase past $x_b$.  Thus
  $[x_a,x_b]$ is a trapping region and there is at least one periodic
  solution.
  
  Linearizing equation (\ref{ge}) shows that a solution $x_0(t)$ is
  stable if
  \begin{equation}
    \label{eq:stab_cond}
    \lambda = \int_0^1 \frac{\partial f}{\partial x}(\hat{x}(s))ds
  \end{equation}
  is negative.  But since $f'(x) < 0$ for all $x \in [x_a,x_b]$, every
  periodic solution is stable.
\end{proof}

\begin{cor}
  Let $x_m$ be such that $f(x_m) = \max_{x>0} f(x)$, where $f(x)$ is
  either $\tanh\beta x - x$ or\, $\gamma x - x^3$.  Then for $h <
  f(x_m)$ equation (\ref{ge}) has exactly three solutions, one in each
  of the intervals $(-\infty, -x_m)$, $(-x_m, x_m)$ and $(x_m,
  \infty)$.  Two of these solutions are stable and one is unstable.
\end{cor}

%%%%%%%%%%%%%%%%%%%%%%% Section %%%%%%%%%%%%%%%%%%%%%%%%%%%%%%%%%%%%%%%%%
\subsection{Uniqueness of periodic solution for large $h$}

For large $h$ we have the following general theorem, which, in
particular, is valid for both functions $f(x)$ that are of interest to us.

\begin{theo} \label{unique_large_h}
  Let $f(x)$ be a continuous almost everywhere differentiable
  function.  If its derivative satisfies
  \begin{eqnarray}
    \label{eq:der_cond}
    f'(x) < \alpha < 0 \qquad \mbox{for } |x| > x_0
    \qquad\mbox{and}\qquad f'(x) < \beta \qquad \mbox{ for all } x
  \end{eqnarray}
  for some $\alpha$, $\beta$ and $x_0$, then, for sufficiently large $h$,
  equation~(\ref{ge}) has a unique stable periodic solution.
\end{theo}

\begin{proof}
  Existence.  Property~(\ref{eq:der_cond}) implies that $f(x) \to \mp
  \infty$ eventually monotonically as $x\to\pm\infty$.  This means, in
  particular, that the equations $f(x) = h$ and $f(x) = -h$ have
  exactly one solution each for sufficiently large $h$.  Denoting these
  solutions by $x_{h-}$ and $x_{h+}$, we notice that the interval
  $[x_{h-}, x_{h+}]$ is trapping for the trajectories of
  (\ref{eq:der_cond}).  Hence there must be at least one periodic
  solution.
  
  Uniqueness.  We will prove the stability of {\em all}\/ periodic
  solutions which will imply that there can be at most one.  Let
  $\hat{x}(t)$ be a periodic solution, which is stable if $\lambda$,
  defined by equation (\ref{eq:stab_cond}), is negative.  Denoting the
  time spent by the solution inside $[-x_0, x_0]$ by $t_+$, we can
  estimate $\lambda$ by
  \begin{equation}
    \label{eq:est_lambda}
    \lambda < \beta t_+ + \alpha (1-t_+).
  \end{equation}
  We aim to show that the time $t_+ \to 0$ as $h\to\infty$ and hence
  $\lambda < 0$.
  
  %=== Sketch of an orbit crossing [-x_0, x_0] ============
  \begin{figure}[t]
    \centerline{\includegraphics[scale=0.8]{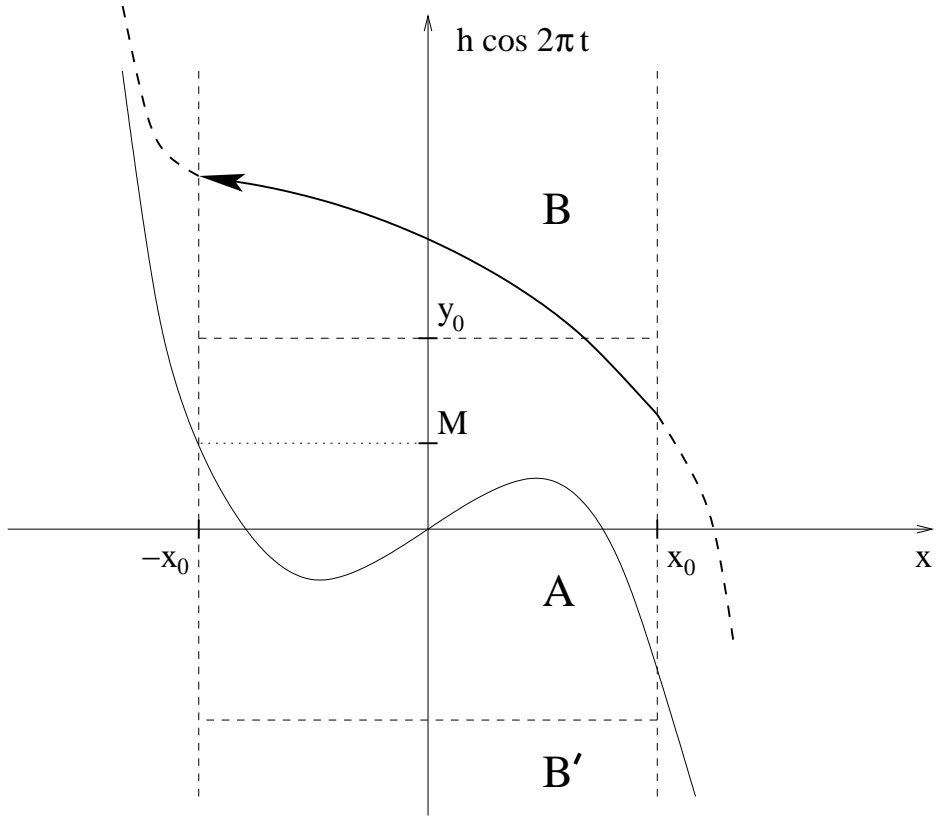}}
    \caption{Sketch of an orbit crossing $[-x_0, x_0]$ and the three
      regions dividing the strip.}
    \label{fig:large_h}
  \end{figure}
  %=========================================================

  To do it we define three regions in the plane $(x, h\cos(2\pi t))$
  which add up to the strip $|x| < x_0$, see figure \ref{fig:large_h}.
  The region $A$ is defined by $|h\cos(2\pi t)| \leq y_0$, the region
  $B$ by $h\cos(2\pi t) > y_0$ and the region $B'$ by $h\cos(2\pi t) <
  -y_0$.  For now, we leave the choice of $y_0$ open.

  From the definition of region $A$, the time spent by the solution in
  $A$ is at most
  \begin{equation}
    \label{eq:t_A}
    t_A = 2\frac{1}{2\pi}\arcsin\left(\frac{y_0}{h}\right).
  \end{equation}
  If $M$ is the maximum of $|f(x)|$ on $[-x_0, x_0]$, then
  $|x'(t)|>y_0-M$ and the time spent by the solution in $B$ or $B'$
  during one visit is at most
  \begin{equation}
    \label{eq:t_B}
    t_B = \frac{2 x_0}{y_0 - M}.
  \end{equation}
  It is easy to see that a periodic solution can visit $B$ at most
  once during one period (and at most twice if it is not periodic);
  the same applies to region $B'$.  Thus we conclude that the total
  time spent in the strip $|x| < x_0$ is
  \begin{equation}
    \label{eq:t_+}
    t_+ < \frac{1}{\pi}\arcsin\left(\frac{y_0}{h}\right) 
    + \frac{4 x_0}{y_0 - M}.
  \end{equation}
  Now if we take $y_0$ to be equal, for example, to $\sqrt{h}$, the
  right hand side will tend to zero as $h\to\infty$.
\end{proof}

%%%%%%%%%%%%%%%%%%%%%%%%%%%%%%%%%%%%%%%%%%%%%%%%%%%%%%%%%%%%%%%%%%%%%%%%%
%%%%%%%%%%%%%%%%%%%%%%% Section %%%%%%%%%%%%%%%%%%%%%%%%%%%%%%%%%%%%%%%%%
\section{The Liapunov-Schmidt reduction} 
\label{LS}

In this section we assume that function $f(y)$ in equation~(\ref{ge})
is odd.  It immediately follows that if $x(t)$ is a solution of
(\ref{ge}), then so is $\widetilde{x}(t) = -x(t+1/2)$.  In particular,
there is always a ``symmetric solution'' which satisfies $x(t) =
-x(t+1/2)$.

We shall perform the Liapunov Schmidt reduction (LSR) for equation
(\ref{ge}) aiming to determine the character of the bifurcation of the
symmetric solution as we change the parameter $h$.  We shall denote by
$h_{cs}$ the corresponding\footnote{the subscripts ``cs'' stand for
  ``critical symmetric''} critical value of $h$.

We start by defining the operator $\Phi$ by
\begin{equation}
  \label{eq:op_Phi}
  \Phi (x) = \dot{x} - f(x) - h\cos(2\pi t).
\end{equation}
The periodic orbits are the zeros of this operator in the space of
continuously differentiable 1-periodic functions $C^1_p([0,1])$ contained
in $L^2([0,1])$.  The reduction, based on a particular solution $x_0(t)$,
leads to the construction of a reduced function $g\colon
\R\times\R\to\R$ such that the solutions of $g(x,h) = 0$ are locally
in one-to-one correspondence with the solutions of (\ref{eq:op_Phi}),
with the solution $x_0(t)$ corresponding to the zero solution of $g(x,h)=0$.
It is rarely possible to compute $g(x,h)$ explicitly, but one can
examine the bifurcation picture by computing the derivatives of $g$.

First of all, however, we describe the symmetric solution $x_0(t)$ at
the critical point $h=h_{cs}$.

%%%%%%%%%%%%%%%%%%%% Subsection %%%%%%%%%%%%%%%%%%%%%%%%%%%%%%%%%%%%%
\subsection{The bifurcation condition}

The necessary bifurcation condition is that the differential of $\Phi$
has non-zero kernel.  This is because if $\Ker(d\Phi(x_0)) = \{0\}$,
the solution $x_0$ of the equation $\Phi(x_0, h_{cs}) = 0$ can be
uniquely continued by the implicit function theorem.  In our case the
differential of $\Phi$, which we denote by $L$, is
\begin{equation}
  \label{eq:dPhi}
  d\Phi(x) (v_1) = \left. \frac{d}{d\xi} \Phi(x + \xi v_1) \right|_{\xi=0}
  = \dot{v}_1 - \frac{\partial f}{\partial x}(x) v_1 \equiv Lv_1
\end{equation}
Therefore, the necessary condition for bifurcation at $x_0\equiv
x_0(h,t)$ is that there is a non-zero 1-periodic solution to
\begin{equation}
  \label{eq:bif_cond2}
  \dot{v} = \frac{\partial f}{\partial x}(x_0) v.
\end{equation}
The general solution of (\ref{eq:bif_cond2}) is
\begin{equation}
  \label{eq:v_soln}
  v(t) = C \exp\left(
    \int_0^t \frac{\partial f}{\partial x}(x_0(s))ds 
  \right).
\end{equation}
This solution is 1-periodic iff 
\begin{equation}
  \label{eq:bif_cond}
  \int_0^1 \frac{\partial f}{\partial x}(x_0(s))ds = 0,
\end{equation}
which we will refer to as the {\em bifurcation condition}.

To proceed with the LSR we need to find a basis for $\Ker L$ and
$(\Range L)^\bot$.  The former is spanned by the function $v(t)$
defined by (\ref{eq:v_soln}) with, for example, $C=1$.  The latter
satisfies
\begin{equation}
  \label{eq:v_star_def}
  0 = \left\langle v^*, Ly \right\rangle 
  = \left\langle -\frac{d}{dt} v^* -
  \frac{\partial f}{\partial x}(x_0(t))v^*, y\right\rangle
\end{equation}
for all $y$.  Solving equation 
\begin{equation}
  \label{eq:v_star_eq}
  \dot{v}^* = -\frac{\partial f}{\partial x}(x_0) v^*
\end{equation}
and again setting the arbitrary constant equal to one, we obtain
\begin{equation}
  \label{eq:v_star_soln}
  v^*(t) = \exp\left(
    - \int_0^t \frac{\partial f}{\partial x}(x_0(s))ds
    \right).
\end{equation}
Note that with our choice of arbitrary constants $v(t) v^*(t) \equiv
1$ which will be used often in what follows.

%%%%%%%%%%%%%%%%%%%%%%%% Subsection %%%%%%%%%%%%%%%%%%%%%%%%%%%%%%%%%%%%%
\subsection{Symmetries of the reduced function}

The function $g(x, h)$ inherits the symmetries of the original
operator $\Phi(x)$.  In our case, the following proposition from
\cite[4.3]{GS} holds

\begin{prop}
  \label{prop:symmetry}
  Let $dim\Ker L = dim\Range L = 1$ and let $\Ker L =
  \mbox{span}\{v\}$.  If there is a symmetry operator $R\colon
  C^1_p([0,1])\to C^1_p([0,1])$ which satisfies
  \begin{equation}
    \label{eq:comm}
    R^2x = x, \qquad \Phi(Rx) = R\Phi(x) 
    \quad\mbox{and}\quad Rx_0(t) = x_0(t),
  \end{equation}
  for all $x\in C^1_p([0,1])$ then $Rv(t)$ is equal to either $v(t)$
  or $-v(t)$.  In the latter case the reduced function is odd in $x$:
  $g(-x, h) = -g(x, h)$.
\end{prop}

It is easy to check that the symmetry operator satisfying conditions
(\ref{eq:comm}) is $R\phi(t) = -\phi(t+1/2)$.  Since $v(t)$ is an
exponential (see (\ref{eq:v_soln})), its sign does not change over
$[0,1]$ and $Rv(t) = -v(t+1/2)$ must be equal to $-v(t)$.  Therefore
our reduced function is indeed odd.

As a consequence, we immediately get
\begin{equation}
  \label{eq:even_drv_of_g}
  g(0, h) = 0, \qquad g_{xx}(0, h) = 0, \qquad \mbox{etc.}
\end{equation}
In addition, $g_x(0,h_{cs}) = 0$ since $h_{cs}$ is critical.  Thus, to
investigate the pitchfork bifurcation of the symmetric solution of
(\ref{ge}), if the bifurcation is indeed a pitchfork, it suffices to
study $g_{xh}(0, h_{cs})$ and $g_{xxx}(0, h_{cs})$.

%%%%%%%%%%%%%%%%%%%%%%%% Subsection %%%%%%%%%%%%%%%%%%%%%%%%%%%%%%%%%%%%%
\subsection{Bifurcation function}

To study the remaining derivatives of $g(x,h)$ we compute higher
derivatives of the operator $\Phi$ (here $v_1, v_2, v_3 \in C^1_p([0,1])$):
\begin{eqnarray}
  \label{eq:higher_dPhi}
  d^2\Phi(x) (v_1, v_2) 
  &=& - \frac{\partial^2 f}{\partial x^2}(x) v_1 v_2\\
  d^3\Phi(x) (v_1, v_2, v_3)
  &=& - \frac{\partial^3 f}{\partial x^3}(x) v_1 v_2 v_3\\
  \frac{\partial}{\partial h}\Phi &=& -\cos(2\pi t)\\
  d\left(\frac{\partial}{\partial h}\Phi\right) &=& 0.
\end{eqnarray}

We have already concluded that
\begin{eqnarray}
  \label{eq:g_h}
  g_h &=& \left\langle v^*,\, \frac{\partial}{\partial h}\Phi
  \right\rangle = 0,\\
  \label{eq:g_xx}
  g_{xx} &=& \Big\langle v^*,\, d^2\Phi(v, v) 
  \Big\rangle = 0.
\end{eqnarray}
These formulae will come in handy below. 

The derivative $g_{hx}$ is given by
\begin{equation}
  \label{eq:g_hx_form}
  g_{hx} = \left\langle v^*, d\left(\frac{\partial}{\partial h}\Phi\right)v
  - d^2\Phi\left(v, L^{-1}E\left(\frac{\partial}{\partial h}\Phi
    \right)\right) \right\rangle,
\end{equation}
where $E$ is the projection to $\Range L$,
\begin{equation}
  \label{eq:def_E}
  Ey = y - \langle v^*, y\rangle \frac{v^*}{\|v^*\|}.
\end{equation}
In our case we have $d\left(\frac{\partial}{\partial h}\Phi\right) =
0$ and, since $\left\langle v^*, \frac{\partial}{\partial
    h}\Phi\right\rangle = g_h = 0$, the projector $E$ leaves
$\frac{\partial}{\partial h}\Phi$ invariant.  Taking into account that
$v^*v \equiv 1$ we get
\begin{equation}
  \label{eq:g_hx_calc}  
  g_{hx} = -\left\langle v^*, 
    d^2\Phi\left(v, L^{-1}\left(\frac{\partial}{\partial h}\Phi
      \right)\right) \right\rangle
  = \int_0^1 \frac{\partial^2 f}{\partial x^2}(x_0(s)) u(s) ds,
\end{equation}
where $u(s)= L^{-1}\left(\frac{\partial}{\partial h}\Phi\right)$ is
the solution of
\begin{equation}
  \label{eq:u_def}
  \dot{u} - \frac{\partial f}{\partial x}(x_0(s)) u = -\cos(2\pi t).
\end{equation}
Differentiating the identity $\Phi(x_0(h,t)) = 0$ with respect to $h$,
we notice that $u(s) = -\partial x_0(h,s)/\partial h$.  Eventually we
get
\begin{equation}
  \label{eq:g_hx_res}
  g_{hx}(0, h_{cs}) = -\frac{\partial }{\partial h} 
  \int_0^1 \frac{\partial f}{\partial x}(x_0(s)) ds,
\end{equation} 
therefore, at the stability-gaining bifurcation we should have
\begin{equation}
  \label{eq:g_hx_conclusion}
  g_{hx} \geq 0.
\end{equation}

\begin{rem}
  One consequence of theorem~\ref{unique_large_h} is that the symmetric
  solution has to undergo {\em at least one} bifurcation in which it
  turns from being unstable to being stable.  As of now, we are unable
  to prove that it undergoes exactly one bifurcation which is
  equivalent to showing that $g_{hx}$ is strictly greater than zero at
  all bifurcations.
\end{rem}

The character of the stability-gaining bifurcation is determined by
the sign of $g_{xxx}$,
\begin{theo} \label{mult1}
  If $g_{xxx}(0, h_{cs})$ is positive (corresp.~negative), the
  stability-gaining bifurcation from the zero-mean solution is a
  subcritical (corresp.~supercritical) pitchfork.
\end{theo}

\begin{proof}
  We expand $g(x, h)$ in Maclaurin series in $x$, taking into account
  that the function is odd,
  \begin{equation}
    \label{eq:g_taylor}
    g(x, h) = g_x(0, h) x + g_{xxx}(0, h) \frac{x^3}{6} + O(x^5)
  \end{equation}
  To understand the bifurcation picture, it suffices to take into
  account only the first two terms.  If $g_{xxx}(0, h) > 0$,
  the equation
  \begin{equation}
    \label{eq:g_taylor_0}
    g_x(0, h) x + g_{xxx}(0, h) \frac{x^3}{6} = 0
  \end{equation}
  has 3 solutions if $g_x(0, h) < 0$ and 1 solution otherwise.  Since
  $g_{xxx}(0, h_{cs})>0$ it remains so in a small neighbourhood of
  $h=h_{cs}$.  Thus the main question is the sign of $g_x(0, h)$.
  
  To prove that, as $h$ crosses $h_{cs}$, $g_x(0, h)$ goes from being
  negative to being positive, we can show that\footnote{of which
    (\ref{eq:g_hx_res}) is a consequence but not a proof}
  \begin{equation}\label{g_xh_h}
    g_x(0, h) = - \int_0^1 \frac{\partial f}{\partial x}(x_0(s)) v(s)
    \, ds = -\lambda
  \end{equation}
  which is rather involved.  Alternatively we can quote a result
  connecting the sign of $g_x(x, h)$ with the stability of the
  corresponding solution of the original differential equation
  \cite{GS}.  In our case, the solution corresponding to $(0, h)$ is
  the symmetric solution which is stable (unstable) whenever $g_x(0,
  h)$ is positive (corresp.~negative).  Since the bifurcation is
  stability-gaining, we get three solutions for $h<h_{cs}$ and one
  solution for $h>h_{cs}$.
\end{proof}

To study the sign of $g_{xxx}$ we use the following formula
\begin{equation}
  \label{eq:g_xxx_formula}
  g_{xxx} = \Big\langle v^*, \ d^3\Phi(v, v, v)
  - 3d^2\Phi\left(v, L^{-1}E d^2\Phi(v,v)\right)\Big\rangle
\end{equation}

From the definition of the projector $E$, see (\ref{eq:def_E}), and identity
(\ref{eq:g_xx}) we conclude that $Ed^2\Phi(v,v) = d^2\Phi(v,v)$.
Further, inverting $L$ we get
\begin{equation}
  \label{eq:L-1}
  \fl  L^{-1} d^2\Phi(v,v) = v(t) \int_0^t \Phi(v,v) v^*(s)\, ds
  =  - v(t) \int_0^t \frac{\partial^2 f}{\partial x^2}(x(s)) v(s)\, ds.
\end{equation}
First we evaluate the second part of (\ref{eq:g_xxx_formula}),
\begin{eqnarray}
  \label{eq:sec_part_gxxx}
  \fl \nonumber
  -3 \Big\langle v^*, \ d^2\Phi\left(v, 
    L^{-1}E d^2\Phi(v,v)\right)\Big\rangle\\
  = -3 \int_0^1 v^*(t) \frac{\partial^2 f}{\partial x^2}(x) v^2(t) 
  \left(\int_0^t \frac{\partial^2 f}{\partial x^2}(x(s)) v(s)\, ds
  \right)\, dt,
\end{eqnarray}
which can be written as
\begin{equation}
  \label{eq:sec_part_gxxx_cont}
  -3 \int_0^1 w'(t) w(t) \, dt = \left[-\frac{3w^2(t)}{2} \right]_0^1 = 0,
\end{equation}
where $w(t) = \int_0^t \frac{\partial^2 f}{\partial x^2}(x(s)) v(s)\,
ds$.  The integral above is equal to zero since $w(1) = g_{xx} = 0$.

Thus, in our case,
\begin{equation}
  \label{eq:g_xxx_res}
  g_{xxx} = \Big\langle v^*, \ d^3\Phi(v, v, v)\Big\rangle
  = -\int_0^1 \frac{\partial^3 f}{\partial x^3}(x_0(s)) v^2(s) ds.
\end{equation}

\begin{cor}
  The stability-gaining bifurcation of the zero-mean solution of
  (\ref{ge}) with $f(x) = x - x^3$ is a subcritical pitchfork.
\end{cor}

\begin{proof}
  In this case $\frac{\partial^3 f}{\partial x^3} \equiv -6$ and,
  since $v^2$ is non-zero, $g_{xxx} > 0$.
\end{proof}

A different proof of this (local) theorem for cubic $f$ is given by
Byatt-Smith \cite{BS}. That in the optical bistability equation (\ref{ob})
one can never have more than three 1-periodic solutions is
proved very elegantly in \cite{Pliss}.

The above results will be used in section \ref{small} to show that locally
the situation in the Suzuki-Kubo equation (\ref{nsk}) is similar to
(\ref{ob}) if $\beta$ is sufficiently small.

%%%%%%%%%%%%%%%%%%%%%%%%%%%%%%%%%%%%%%%%%%%%%%%%%%%%%%%%%%%%%%%%%%%%%%%%%
%%%%%%%%%%%%%%%%%%%%%%% Section %%%%%%%%%%%%%%%%%%%%%%%%%%%%%%%%%%%%%%%%%
\section{The Suzuki-Kubo equation} \label{inf}

In this section we study the case
\begin{equation}
  \label{eq:SK_f}
  f(x) = x - \tanh(\beta x)
\end{equation}
in detail.

We aim to demonstrate that for sufficiently large $\beta$ there are
values of $h$ for which there exist at least $5$ periodic solutions of
(\ref{nsk}).  We also show that for small $\beta$ the bifurcation of
the zero-mean solution is a subcritical pitchfork (figure
\ref{fig:fork}(a)) which means that the maximum number of solutions
is 3.  The boundary between these two regimes is known as the
tricritical point (TCP).

%%%%%%%%%%%%%%%%%%%%%%%% Subsection %%%%%%%%%%%%%%%%%%%%%%%%%%%%%%%%%%%%%
\subsection{The case of small $\beta$} \label{small}

\begin{theo}\label{theo:small_beta}
  There exists $\beta_0$ such that for all $1 < \beta < \beta_0$ and the
  zero-mean periodic solution $x_0(t)$ satisfying~(\ref{eq:bif_cond})
  we have
  \begin{equation}
    \label{eq:diff3_positive}
    \int_0^1 \frac{\partial^3 f}{\partial x^3}(x_0(s)) v^2(s) \ ds < 0.
  \end{equation}
\end{theo}

It is not hard to get the rough estimate, $\beta_0=4/3$, which is sufficient
for our purposes.

\begin{proof}
  Put $\tanh(\beta x_0(s))=\zeta(s)$.  From (\ref{eq:bif_cond2}), we have
  \begin{equation} \label{vn}
    \int_0^1 \frac{\partial f}{\partial x} (x_0(s)) v^n (s)\,ds 
    = \int_0^1 v^{n-1}(s) \dot{v}(s)\,ds = 0
    \quad\hbox{for all } n \geq 1,
  \end{equation}
  since $v(t)$ is 1-periodic.  Differentiating~(\ref{eq:SK_f}),
  \begin{equation}\label{eq:SK_f_der}
    \frac{\partial f}{\partial x} (x_0(s)) = \beta - 1 - \beta \zeta^2(s), 
  \end{equation}
  multiplying it by $v^2(s)$ and choosing $n=2$ in~(\ref{vn}), we have
  \begin{equation} \label{est1}
    \int_0^1 \zeta^2(s) v^2(s) \, ds 
    = \frac{\beta - 1}{\beta} \int_0^1 v^2(s) ds.
  \end{equation}
  
  For the third derivative of $f(x)$ we now get
  \begin{eqnarray}
    \fl\nonumber
    \int_0^1 \frac{\partial^3 f}{\partial x^3}  (x_0(s)) v^2(s)\, ds\\
    \nonumber
    = -2\beta^3 \int_0^1 v^2(s)\, ds 
    + 8\beta^3 \int_0^1 v^2(s)\zeta^2(s) \, ds
    - 6 \beta^3 \int_0^1 \zeta^4(s) v^2(s) \zeta\, ds\\ 
    \nonumber
    \leq -2\beta^3 \int_0^1  v^2(s) \, ds 
    + 8\beta^3 \int_0^1 \zeta^2(s) v^2(s) \, ds\\
    \label{3rd}
    = 2\beta^2(3\beta-4) \int_0^1 v^2(s) \, ds,
  \end{eqnarray}
  where we discarded the integral of a non-negative function and then
  used (\ref{est1}).  Thus the integral on the left-hand side of
  (\ref{3rd}) is negative if $\beta < 4/3$.
\end{proof}

\begin{rem}
  It is easy to improve the above estimate to $3/2$ if, instead of
  discarding the integral of $\zeta^4v^2$ altogether, we deduce from
  (\ref{eq:SK_f_der})
  \begin{equation}
    \label{eq:z4_ineq}
    \zeta^4 = \left(\frac{\beta-1}{\beta} 
    - \frac{1}{\beta} \frac{\partial f}{\partial x}(x_0)\right)^2 
    \geq \left(\frac{\beta-1}{\beta}\right)^2 
    - \frac{2}{\beta}\frac{\partial f}{\partial x} (x_0).
  \end{equation}
\end{rem}

%%%%%%%%%%%%%%%%%%%%%%%% Subsection %%%%%%%%%%%%%%%%%%%%%%%%%%%%%%%%%%%%%
\subsection{The $\beta= \infty$ case} 

In this section all assertions are verified by explicit computation.

When we take $ \beta=\infty$, instead of the ODE (\ref{nsk}) we are
left with the differential inclusion
\begin{equation} 
  \label{eq:inclusion}
  \epsilon \frac{dx}{dt} \in -x +\hbox{Sgn}(x)+ h \cos(2\pi t),
\end{equation}
where $\hbox{Sgn}(x)$ is the multivalued sign function,
\begin{equation}
  \label{eq:Sgn}
  \hbox{Sgn}(x) = 
  \cases{
    -1, & if $x < 0$,\\
    [-1,1], & if $x = 0$,\\
    1, & if $x = 1$.
  }
\end{equation}

Note that if a periodic solution $x(t)$ has no internal zeroes, it solves a
simple linear ODE and can be computed explicitly. For example, a for $h$
sufficiently small, a positive periodic solution is given by
\begin{equation}
  \label{eq:right_soln}
  x(t) = 1 + h\frac{\cos(2\pi t) + 2\pi\epsilon \sin (2\pi t)}
  {1+4\pi^2\epsilon^2},
\end{equation}
from which we conclude that this (asymmetric) solution touches zero when 
\begin{equation}\label{hca}
  h = h_{ca} \equiv \sqrt{1 + 4\pi^2\epsilon^2}.
\end{equation}
For example, if $\epsilon=0.05$, $h_{ca}= 1.048187027$ to 10 d.p.

%=== Central soln sketch =================================
\begin{figure}[t]
  \centerline{
    \includegraphics[scale=0.75]{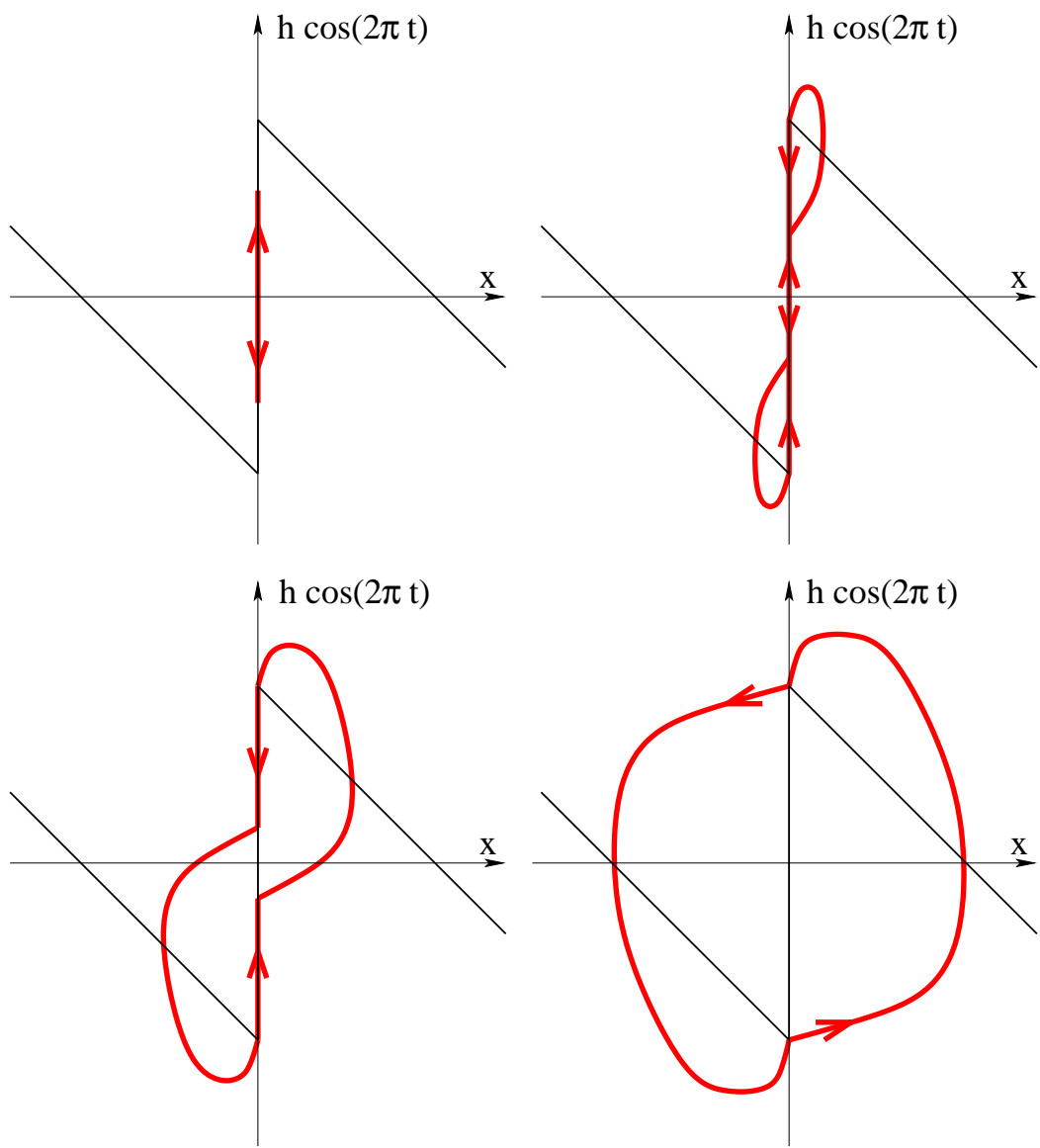}
  }
  \caption{Evolution of the unstable zero-mean solution of 
    inclusion~(\ref{eq:inclusion}) for $h \leq h_{cs}$.  Top left:
    solution is identically zero when $h<1$.  Top right and bottom
    left: emergence and growth of nontrivial parts.  The solution will
    experience a bifurcation when the trivial parts, where $x(t)=0$ on
    an open interval, disappear completely.  This happns at $h=h_{cs}$
    (bottom right).}
  \label{fig:bi_A}
\end{figure}
%=========================================================

Next we want to understand the bifurcation\footnote{a complete
  bifurcation picture and animated periodic solutions described in
  this section are also available at
  \url{http://www.math.tamu.edu/~berko/bistable}} of the zero-mean
solution, figure~\ref{fig:bi_A}.  When $h < 1$, the zero-mean solution
is identically zero and is unstable.  After $h$ crosses 1, the
solution gets two symmetric non-trivial parts, but there are still two
intervals in $t$ on which $x(t)$ is identically zero.  The bifurcation
happens when each of these intervals collapses to a point.

Let $h_{cs}$ be the unique value of $h$ for which there exists a
periodic solution $x_c(t)$ of the differential inclusion with the
property that there is $t_0$ such that $x_c'(t_0)=1$ and
$x_c'(t_0+1/2)=-1$, see figure \ref{fig:bi_A}(bottom right).  

%=== Central soln sketch =================================
\begin{figure}[t]
  \centerline{
    \includegraphics[scale=0.75]{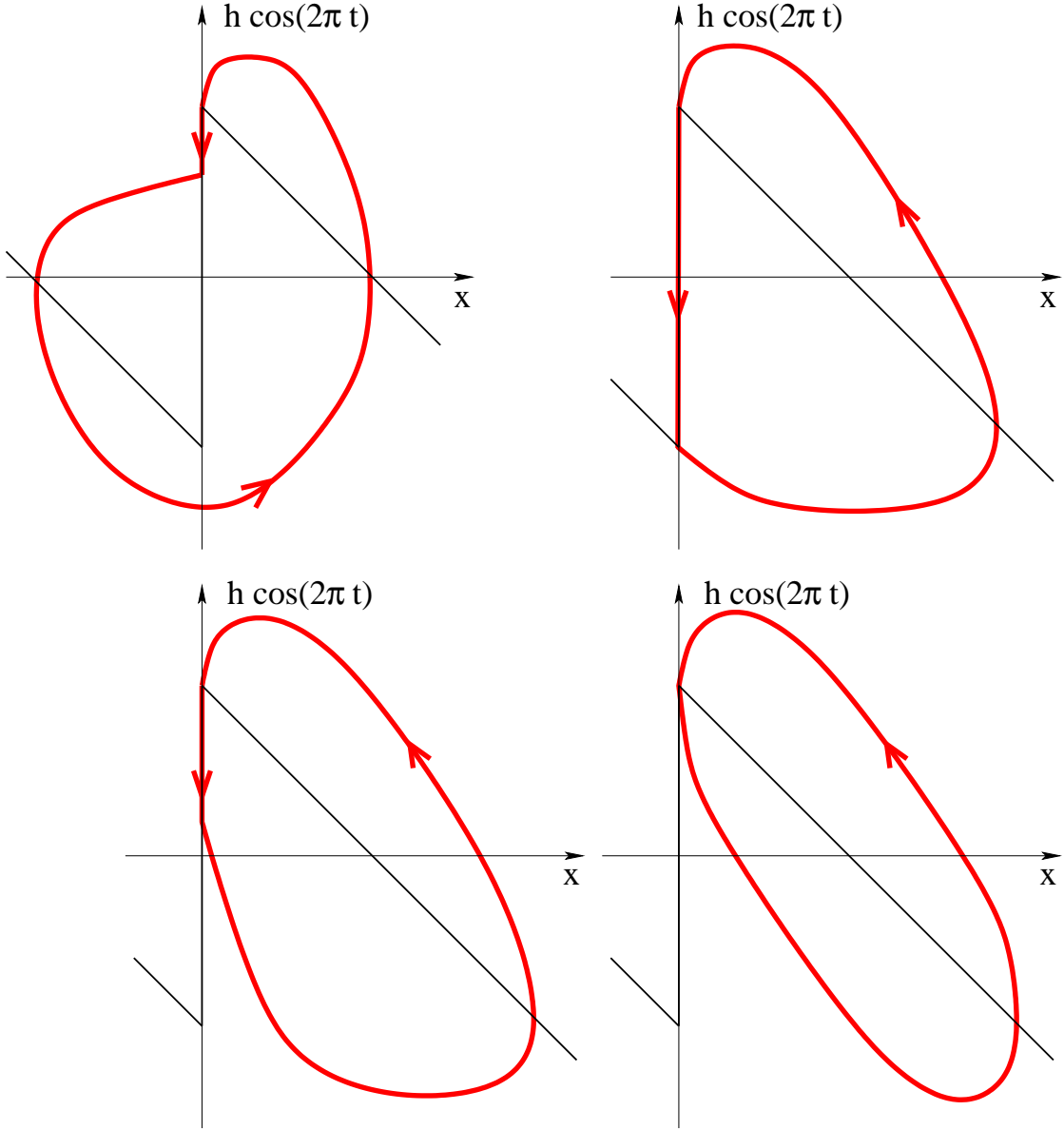}
  }
  \caption{The evolution of the right unstable solution until its 
    annihilation at $h=h_{ca}$.}
  \label{fig:bi_B}
\end{figure}
%=========================================================

An involved but elementary computation (see \ref{app:calc_h_cs}) shows
that this happens for
\begin{equation}\label{hcs}
  h = h_{cs} \equiv \frac{\sqrt{(1+4\pi^2 \epsilon^2)
      (\pi^2 \epsilon^2 -2\pi^2 \epsilon^2 z+(1+\pi^2\epsilon^2)z^2)}}
  {\pi(1+z)\epsilon},
\end{equation}
where we have put $z=\exp(-1/2\epsilon)$.  Expanding this expression in
$z$, we obtain the following expression:
\begin{equation}
  \label{eq:hcs_expansion}
  h_{cs} = h_{ca} - 2h_{ca}z + O(z^2),
\end{equation}
from which it is clear that for small $\epsilon$ (and hence for very small
$z$) $h_{cs}< h_{ca}$. For example, if $\epsilon=0.05$, $h_{cs}=1.048091900$
and therefore $h_{ca}-h_{cs} = O(10^{-4})$.

When $h = h_{cs}$, the zero mean solution becomes stable through a
super-critical pitchfork bifurcation.  It emits a couple of unstable
solutions which disappear in a blue-sky catastrophe, i.e. in a
collision with the stable solutions (\ref{eq:right_soln}).  These
unstable solutions are sketched in Figs.~\ref{fig:bi_B} for various
values of $h$, $h_{cs} < h \leq h_{ca}$.

We summarise the discussion of the inclusion (\ref{eq:inclusion}) in
a theorem

\begin{theo}\label{multinc}
  For $h$ satisfying $h_{cs} < h < h_{ca}$, the inclusion
  (\ref{eq:inclusion}) has five 1-periodic solutions.
\end{theo}

%%%%%%%%%%%%%%%%%%%%%%%%% Subsection %%%%%%%%%%%%%%%%%%%%%%%%%%%%%%%%%%%%
\subsection{The case of large $\beta$}

%=== Three STABLE solutions =================================
\begin{figure}[t]
  \centerline{
    \includegraphics[height=8cm,width=12cm]{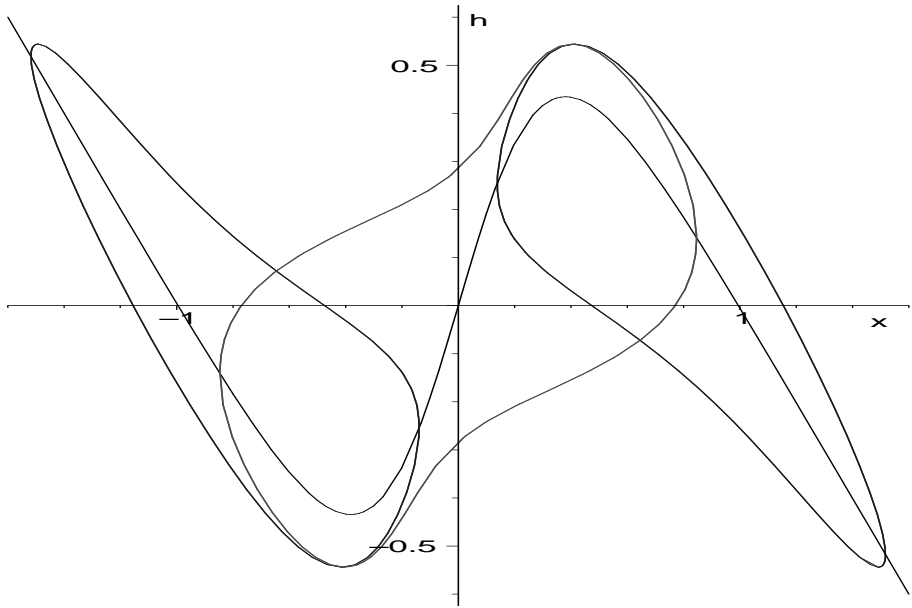}
  }
  \caption{Three coexisting stable solutions of the Suzuki-Kubo
    equation for $\beta = 3$, $\epsilon = 0.05$ and $h = 0.544$.}
  \label{fig:tanh_3stable}
\end{figure}
%=========================================================

It is possible to extend the existence of 5 solutions from
$\beta=\infty$ case to the case of large $\beta$ by continuity.
In our numerical experiments we found that $\beta$ does not have to be
very large.  Figure~\ref{fig:tanh_3stable} contains the plot of 3
stable solutions of equation~(\ref{nsk}) with $\beta = 3$, $\epsilon =
0.05$, $h = 0.544$ and initial conditions $x(0) = -0.413600598,
-0.40456237, 1.491606074$.

\begin{theo} \label{mult2}
  For any $h$ satisfying $h_{cs} < h < h_{ca}$ and sufficiently large
  (depending on $h$) $\beta$ equation~(\ref{nsk}) has at least five
  periodic solutions.
\end{theo}

The proof of this theorem is based on an upper semicontinuity result
for differential inclusions \cite{AC}. To state it, we need to
introduce some notation. Let $B$ be the unit ball in $\R^n$ with
centre at the origin.  Consider a differential inclusion
\begin{equation} \label{gde}
  x'(t) \in F(t, x(t), \lambda), 
\end{equation}
$(t, x, \lambda) \in [0,T] \times \Omega \times \Lambda$, where
$\Omega$ is a bounded subset of $\R^n$, $\Lambda\subset\R$ and $F$ is
an upper semicontinuous set-valued map with uniformly bounded compact
images.  Assume that for all $\lambda\in\Lambda$ and any $x_0$ in some
set $Q\in\Omega$ the solutions of the differential inclusion with
$x(0)=x_0$ exist on the interval $[0,T]$ and remain in $\Omega$.  The
Poincar\'e map ${\cal P}_\lambda^T(x_0)\colon Q\to\R^n$ is the
set-valued map defined by
\begin{equation}
  \label{eq:poincare}
  {\cal P}_\lambda^T(x_0) 
  = \{x(T) \colon x'(t) \in F(t, x(t), \lambda), x(0) = x_0\}.
\end{equation}
The following theorem is a slight variation of the continuity theorems
in Section 2.2 of \cite{AC}.

\begin{theo}[Dependence on Parameter]\label{ac}
  Let the map $(t,x,\lambda) \rightarrow F(t,x,\lambda)$ satisfy the
  conditions of the previous paragraph.  Then for any
  $\lambda_0\in\Lambda$ and any open set $U\subset \R^{2n}$, $0\in U$,
  there is a $\delta > 0$ such that $|\lambda-\lambda_0|< \delta$
  implies
  \begin{equation}
    \hbox{graph}  ({\cal P}_\lambda^T) \subset
    \hbox{graph}  ({\cal P}_{\lambda_0}^T) + U.
  \end{equation}
\end{theo}

\begin{proof}[Proof of Theorem \ref{mult2}]
  Clearly, $F(t,x,\beta)$ defined by the right-hand side (\ref{nsk})
  viewed as a differential inclusion satisfies the conditions of
  Theorem~\ref{ac} since all solutions of this inclusion are globally
  defined and remain bounded.
  
  Fix $h \in (h_{cs}, h_{ca})$.  By theorem \ref{ac}, for sufficiently
  large $\beta$ the graph of the Poincar\'e map ${\cal P}_\beta^1$ of
  the Suzuki-Kubo equation (\ref{nsk}) is contained in a neighbourhood
  of the graph of ${\cal P}_\infty$.  By Theorem~\ref{multinc}, the
  graph has five intersections with the diagonal.  Therefore, for
  sufficiently small neighbourhood, i.e.~for sufficiently large
  $\beta$, we get at least five $1$-periodic solutions of (\ref{nsk}).
\end{proof}

%%%%%%%%%%%%%%%%%%%%%%%%%%%%%%%%%%%%%%%%%%%%%%%%%%%%%%%%%%%%%%%%%%%%%%%%%
%%%%%%%%%%%%%%%%%%%%%%% Section %%%%%%%%%%%%%%%%%%%%%%%%%%%%%%%%%%%%%%%%%

\section{Conclusions}

In this paper we have resolved a minor controversy in the physical
literature. However, some major questions still remain. One we can
formulate as a conjecture:

\begin{conj} 
  An ODE of the form of (\ref{ge}) with $f(y)$ bistable has
  at most five 1-periodic solutions.
\end{conj}

The other fundamental question is to formulate necessary conditions on
$f(y)$ for (\ref{ge}) to have a specified maximum number of solutions.  
We end with two remarks.
\begin{enumerate}
\item 
  Consider (\ref{ge}) in which the nonlinearity $f(y)$ is piecewise linear,
  \begin{equation}
    f(y) = 
    \cases{
      -1-y & if $y < -1/2$ \\
      y  & if $y \in [-1/2, 1/2]$\\
      1-y & if $y > 1/2$.
    }
  \end{equation}
  Then it can be shown, by explicitly constructing the stable periodic
  solutions, that there is a range of $h$ for which (\ref{ge}) with such
  a nonlinearity will have five 1-periodic solutions\cite{Berk}.  This
  perhaps indicates that there may be more than one mechanism for the
  appearance and disappearance of solutions.
\item
  A possible way to attack the problem of necessary conditions is to
  consider a homotopy of (\ref{ob}) into (\ref{sk}), i.e.
  \begin{equation}
    x' = (1-2\alpha)x + (1-\alpha)\tanh (\beta x) 
    - \alpha x^3 + h\cos(2\pi t),
  \end{equation}
  and see, at least numerically, for which value $\alpha$ one no longer has
  a subcritical bifurcation from the symmetric solution. 
\end{enumerate}

\section*{Acknowledgement}

The authors gratefully acknowledge fruitful discussions with
J.~Byatt-Smith, B.~Duffy and J.~Carr.

\appendix

%%%%%%%%%%%%%%%%%%%%%%%%%%%%%%%%%%%%%%%%%%%%%%%%%%%%%%%%%%%%%%%%%%%%%%%%%
%%%%%%%%%%%%%%%%%%%%%%%%% Appendix %%%%%%%%%%%%%%%%%%%%%%%%%%%%%%%%%%%%%%
\section{Calculating $h_{cs}$}
\label{app:calc_h_cs}

The definition of $h_{cs}$ is the value of $h$ for which there exists
a periodic solution $x_c(t)$ of equation
\begin{equation}
  \label{eq:lin_right}
  \epsilon \frac{dx}{dt}= 1 - x + h\cos(2\pi t),
\end{equation}
with the property that it connects the points $(0, -1)$ and $(0, 1)$
(see diagram).  In other words, the following equations are satisfied
\begin{eqnarray}
  \label{eq:h_cs_condition1}
  h \cos(2\pi t_0) &=& 1,\\
  \label{eq:h_cs_condition2}
  x_c(t_0) &=& 0,\\
  \label{eq:h_cs_condition3}
  x_c(t_0 + 1/2) &=& 0
\end{eqnarray}

The general solution of equation (\ref{eq:lin_right}) is
\begin{equation}
  \label{eq:lin_right_gen_soln}
  x(t) = C_1e^{-t/\epsilon} 
  + \frac{1+4\pi^2\epsilon^2+h\cos(2\pi t) + 2h\pi\epsilon\sin(2\pi
    t)}{1+4\pi^2\epsilon^2}
\end{equation}
From~(\ref{eq:h_cs_condition1}) we get $h\sin(2\pi t_0)=\sqrt{h^2-1}$.
Equations~(\ref{eq:h_cs_condition2})-(\ref{eq:h_cs_condition3}) then
read
\begin{eqnarray}
  \label{eq:h_cs_condition3_expanded}
  C_1e^{-t_0/\epsilon}
  + \frac{2 + 4\pi^2\epsilon^2
    + 2\pi\epsilon\sqrt{h^2-1}}{1+4\pi^2\epsilon^2} &=& 0\\
  C_1e^{-t_0/\epsilon} e^{-1/2\epsilon}
  + \frac{4\pi^2\epsilon^2 
    - 2\pi\epsilon\sqrt{h^2-1}}{1+4\pi^2\epsilon^2} &=& 0
\end{eqnarray}
We eliminate $C_1e^{-t/\epsilon}$ to get
\begin{equation}
  \label{eq:h_cs_almost}
  e^{-1/2\epsilon} \left(2 + 4\pi^2\epsilon^2
    + 2\pi\epsilon\sqrt{h^2-1}\right)
  - 4\pi^2\epsilon^2 + 2\pi\epsilon\sqrt{h^2-1} = 0
\end{equation}
Denoting $e^{-1/2\epsilon} = z$, we finally obtain
\begin{equation}
  \label{eq:h_cs_answer}
  h_{cs} = \frac{\sqrt{(1 + 4\pi^2\epsilon^2)(\pi^2\epsilon^2 -
      2\pi^2\epsilon^2z + (1 + \pi^2\epsilon^2)z^2)}}{\pi\epsilon(1+z)}
\end{equation}

%%%%%%%%%%%%%%%%%%%%%%%%%%%%%%%%%%%%%%%%%%%%%%%%%%%%%%%%%%%%%%%%%%%%%%%%
%%%%%%%%%%%%%%%%%%%%%%%% References %%%%%%%%%%%%%%%%%%%%%%%%%%%%%%%%%%%%

\end{document}